\documentclass[twocolumn]{revtex4}
\usepackage{color}
\usepackage{graphics,graphicx,epsfig}
\usepackage{epsf,epstopdf,wrapfig}
\usepackage{amssymb,amsfonts,amsmath}
\usepackage[export]{adjustbox}
\include{epsf}
\usepackage{ifthen}		

\newcommand{\beqn}{\begin{eqnarray}}
\newcommand{\eeqn}{\end{eqnarray}}
\newcommand{\beq}{\begin{equation}}
\newcommand{\eeq}{\end{equation}}

\usepackage{ifthen}

\newcommand{\YE}[1]{{\color{black}#1}}

\begin{document}
\title{Predicting the spectrum of TCR repertoire sharing \\
with a data-driven model of recombination}

\author{Yuval Elhanati$^{1\dagger}$, Zachary Sethna$^{1\dagger}$, Curtis G. Callan Jr.$^{1}$,  Thierry Mora$^{2*}$, Aleksandra M. Walczak$^{3*}$}

\affiliation{
	\normalsize{$^{1}$Joseph Henry Laboratories, Princeton University, Princeton, New Jersey 08544 USA}\\
	\normalsize{$^{2}$ Laboratoire de physique statistique, CNRS,
          Sorbonne Universit\'e, Universit\'e Paris-Diderot,
          and \'Ecole Normale Sup\'erieure (PSL University), 24, rue Lhomond, 75005 Paris, France}\\
	\normalsize{$^{3}$Laboratoire de physique th\'eorique, CNRS,
          Sorbonne Universit\'e, and \'Ecole Normale Sup\'erieure (PSL University), 24, rue Lhomond, 75005 Paris, France}\\
\normalsize{${}^{\dagger}$ Equal contribution.}\\
\normalsize{${}^*$ Equal contribution.}\\
	}

\begin{abstract}
Despite the extreme diversity of T cell repertoires, many identical T
cell receptor (TCR) sequences are found in a large number of
individual mice and humans. These widely-shared sequences, often
referred to as \lq public\rq, have been suggested to be
over-represented due to their potential immune functionality or their
ease of generation by V(D)J recombination. Here we show that even for
large cohorts the observed degree of sharing of TCR sequences between
individuals is well predicted by a model accounting for the known quantitative
statistical biases in the generation process, together with a
simple model of thymic selection. Whether a sequence is
shared by many individuals is predicted to depend on the number of queried
individuals and the sampling depth, as well as on the sequence itself,
in agreement with the data. 
We introduce the {\it degree of
  publicness} conditional on the queried cohort size and the size of
the sampled repertoires. Based on these
observations we propose a public/private sequence classifier,
`PUBLIC' (Public Universal Binary Likelihood Inference Classifier), based on the generation probability, which performs very well even for small cohort sizes. 

\end{abstract}

\maketitle

\section{Introduction}

The adaptive immune system relies on a diverse set of T-cell receptors (TCR) to recognize pathogen-derived peptides presented by the major histocompatibility complex (MHC). Each T cell expresses a distinct TCR that is created stochastically by V(D)J recombination. This process is very diverse, with the potential to generate up to $10^{61}$ different sequences in humans \cite{Mora2016}. The resulting `repertoire' of distinct TCRs expressed in an individual defines a unique footprint of immune protection. Despite this diversity, a significant overlap in the TCR response of different individuals to a variety of antigens and infections has been observed in humans \cite{Moss1991,Casanova1992,Argaet1994}, mice \cite{Cibotti1994,Bousso1998,Venturi2006}, and macaques \cite{Venturi2008a} (reviewed in Refs.~\cite{Venturi2008,Li2012}). This observation led to the notion of a `public' response shared by all, and a complementary `private' response specific to each individual \cite{Cibotti1994}. Since antigen-specific TCRs have a restricted set of sequences\cite{Dash2017,Glanville2017}, and since there is no identified analog for T cells of B cell affinity maturation, a public response can only arise if the specific responding T cells are independently generated in each individual's T cell repertoire. 
It was proposed \cite{Venturi2006,Venturi2008a,Venturi2008} that these shared sequences can be explained by the biases inherent in the V(D)J recombination process, together with `convergent recombination', the possibility to generate the same TCR sequence (especially the same CDR3 amino acid sequence) in independent recombination events. In this hypothesis, shared TCRs are simply those that have a higher-than-average generation probability and are thus more abundant in the unselected repertoire \cite{Quigley2010}. The advent of high-throughput sequencing of TCR repertoires \cite{Robins2009,Freeman:2009fja,Benichou2012,Six2013} has largely confirmed this view through the analysis of shared TCR sequences between unrelated humans \cite{Robins2010b,Venturi2011a,Elhanati2014}, monozygous human twins \cite{Zvyagin2014a,Pogorelyy2017a}, and mice \cite{Madi2014}. 
However, despite recent efforts to characterize the landscape of public TCRs \cite{Madi2017}, the relative contributions of convergent recombination, V(D)J bias, thymic selection \cite{Furmanski2008}, peripheral and antigen-specific selection, remain to be elucidated and quantified.

In this review, we address the sharing phenomenon using quantitative models of the stochastic V(D)J recombination process that have been inferred from repertoire data \cite{Murugan2012, Elhanati2016, Sethna2017,Marcou2018}. These generative models, augmented by a simple one-parameter model of thymic selection, can be used to predict the number of sequences that will be shared between any number of individuals, each sampled to any sequencing depth. We make these predictions on the basis of stochastic simulations, but we also derive general mathematical formulas that allow us to calculate sharing from any recombination model. We show that these predictions are in excellent quantitative agreement with data from two recent T cell repertoire studies in humans \cite{Emerson2017} and mice \cite{Madi2014}. Our results are consistent with arguments \cite{Venturi2008,Venturi2013} that the dichotomy between public and private is misleading. Instead, we find a wide range of possible degrees of sharing, depending on sequencing depth of the individual repertoires, the number of individuals in the study, and the number of individuals between whom the sequence is shared. We propose `PUBLIC' (Public Universal Binary Likelihood Inference Classifier), a `publicness score' defined as the recombination probability predicted by our model. This score predicts the sharing status of any TCR with very high accuracy, irrespective of the definition for being public versus private.

\begin{figure*}
\begin{center}
\includegraphics[width=.8\linewidth]{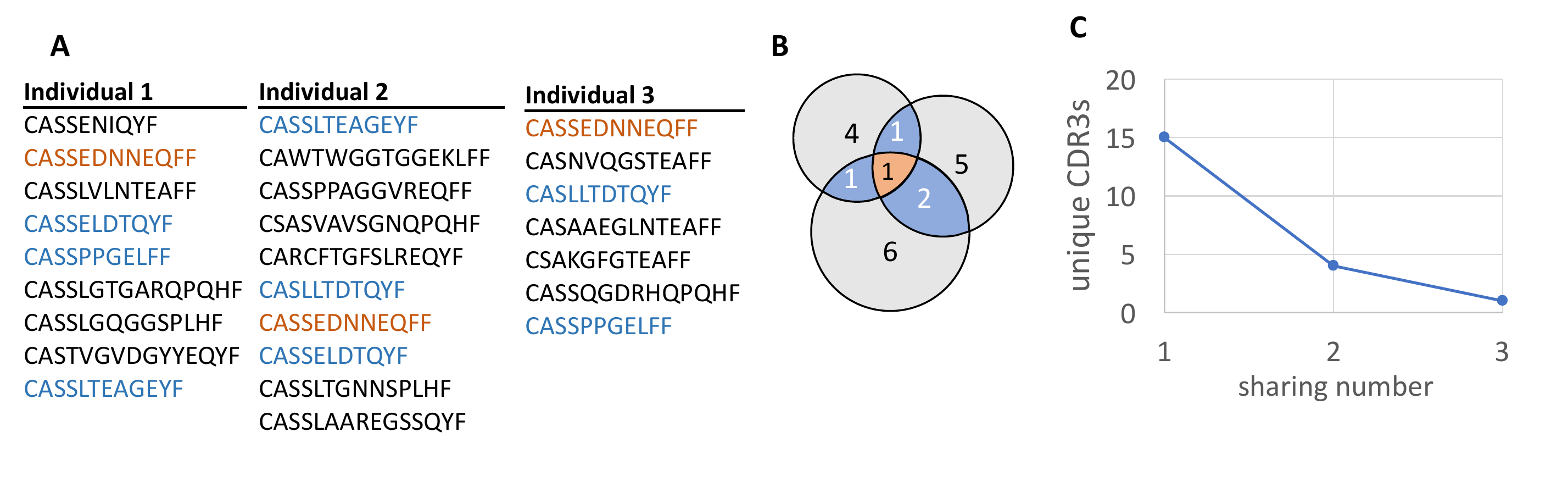}
\caption{A cartoon representation of the pipeline for computing the distribution of shared sequences between samples. (A) Sharing between samples is analyzed by marking repeated CDR3s between $K$ samples. (B) The overlapping sequences are counted and binned, and the number of CDR3s that were shared $m$ times is computed. (C) Distribution of the number of sequences that are shared $m$ times between  the sample of $K$ individuals.}
\label{cartoon}
\end{center}
\end{figure*}

\section{Predicting sharing between repertoires}
\subsection{Spectrum of sharing numbers}
We start with an operational definition of sharing in repertoire data obtained by high-throughput sequencing from several individuals or cell subsets, which closely follows that of Ref.~\cite{Madi2014}. For each individual, we compile a list of unique TCR sequences (Fig.~\ref{cartoon}A). 
Since the functional character of a T cell is thought to be largely determined by the amino acid sequence of the highly variable Complementary Determining Region 3, or CDR3 (to be more precisely defined later) of the beta chain protein, we record in our list just the unique CDR3 beta chain amino acid sequences found in a given biological sample of T cells. For each TCR amino acid sequence, we define the \lq sharing number\rq\, as the number of different samples in which that sequence was found (Fig.~\ref{cartoon}B). The sharing number depends both on the number of samples and on the number of unique sequences in each sample.  We note that more restricted definitions of sharing, based for example on the full nucleotide sequence, are possible, but the correspondingly reduced statistics make it harder to draw sharp conclusions.
Counting the number of TCRs with each sharing number (Fig.~\ref{cartoon}C), we obtain a distribution of sharing, from purely private sequences (sharing number 1) to fully public sequences (sharing number equal to the number of individuals), and everything in between. 
We will compare the distribution of sharing numbers obtained from the data sequences with predictions of our models.

Early estimates of sharing of human TCRs \cite{Venturi2006} showed that assuming a uniform distribution of TCR generation underestimates observed sharing by several orders of magnitude \cite{Robins2010b}. Thus, having an accurate model for the non-uniform distribution of TCR generation probabilities is crucial for making quantitative predictions of the sharing distribution. A simple non-homogeneous model that assigns lower probability to TCR sequences with more N insertions in the V(D)J recombination process is able to predict sharing between pairs of individuals within the correct order of magnitude \cite{Robins2010b}. However, this estimate ignores the detailed structure of biases inherent to the recombination process and results in strong biases in the distribution of TCR sequences that, as we will show, influence the sharing spectrum. 

\subsection{TCR generation bias}
T-cell receptors are composed of an $\alpha$ and a $\beta$ chain encoded by separate genes stochastically generated by the V(D)J recombination process \cite{Hozumi1976}. Each chain is assembled from the combinatorial concatenation of two or three segments (V as Variable, D as Diversity, and J as Joining for the $\beta$ chain, and V and J for the $\alpha$ chain) picked at random from a list of germline template genes. Further diversity comes from random nontemplated N insertions between, together with random deletions from the ends of, the joined segments. The $\alpha$ chain is less diverse than the $\beta$ chain and sharing analyses have mostly focussed on the latter. The germline gene usages are highly non-uniform \cite{Roldan1995,Robins2009,Freeman:2009fja}, due to differences in gene copy numbers \cite{Luo2016} as well as the conformation \cite{Ndifon2012} and processive excision dynamics \cite{Warmflash:2006tu} of DNA during recombination. In addition, the distributions of the number of deleted and inserted base pairs, as well as the composition of N nucleotides, are also biased \cite{Gauss:1996vx}. Taken together, the biases imply that some recombination events are more likely than others. In addition, distinct recombination events can lead to the same nucleotide sequence, and many nucleotide sequences can lead to the same amino-acid sequence. This convergent recombination further skews the distribution of TCRs, as some sequences can be produced in more ways than others \cite{Venturi2006,Venturi2008}.

The effects of recombination biases and convergent recombination can be captured by stochastic models of recombination. Given the probability distributions for the choice of gene segments, deletion profiles and insertion patterns, one can generate  {\em in silico} TCR repertoire samples that mimic the statistics of real repertoires, and allow us to predict sharing statistics and the effects of convergent recombination \cite{Murugan2012,Greenaway2013,Madi2014,Elhanati2014,Pogorelyy2017a,Dash2017}. To obtain accurate predictions, the distributions of recombination events used in the model must closely match repertoire data. This task is made difficult by the fact that, as a consequence of convergent recombination, the specific recombination event behind an observed sequence is not directly accessible. However, methods of statistical inference can be used to overcome this problem and learn accurate models of V(D)J recombination \cite{Murugan2012,Ralph2016,Elhanati2016,Marcou2018}, models which can in turn be used to predict sharing properties of sampled repertoires or of individual TCR sequences. These models have been shown to vary little between individuals, with small differences only in the germline gene usage and remarkable reproducibility in the insertion and deletion profiles \cite{Murugan}. In our analysis we will assume a universal model, independent of the individual.

\subsection{Using TCR recombination models to predict sharing}
We used the above-described models of recombination to predict the distribution of sharing among cohorts of humans and mice. Specifically, we re-analyzed published TCR $\beta$-chain nucleotide sequences of 14 Black-6 mice \cite{Madi2014} and 658 human donors \cite{Emerson2017} (Methods). Individual samples comprised 20,000-50,000 unique sequences for mice, and up to 400,000 for humans. Sequences were translated into amino-acid sequences, and trimmed to keep only the CDR3 loop, defined as the sequence between the last cysteine in the V gene and the first phenylalanine in the J gene \cite{Lefranc2009}. The sharing number of each observed CDR3 amino acid sequence, and the sharing number distribution, were then computed from the data. We chose to focus on the CDR3 amino-acid sequences to get higher sharing numbers than would have been obtained for untrimmed nucleotide sequences, limiting the effects of sequencing errors and allowing for a better comparison to the model.

To obtain model predictions for humans, we used a previously described model for TCR$\beta$ sequence generation inferred by the software package IGoR \cite{Marcou2018} from repertoire data of a single individual \cite{Emerson2017} (Methods). The mouse model was inferred using IGoR from the repertoire data of the 14 animals of \cite{Madi2014}. In both cases, the model is learned from unproductive rearrangements (i.e. with a frameshift in the CDR3) since those sequences give us access to the raw result of recombination, without subsequent effects of selection \cite{Murugan2012}. These unproductive sequences are only used to infer a generative model and are not used in the sharing analysis. A productive (in frame) sequence that is generated in a V(D)J recombination event will not necessarily survive thymic selection to become a functional T cell in the periphery. To model this effect, we assume that there is a probability $q$, independent of the actual sequence but dependent on the species under study, that any given generated sequence will survive thymic selection \cite{Pogorelyy2017}. 
Model sharing predictions are then obtained in two ways: (i) by simulating sequences and selecting them at random with probability $q$ to generate samples of the same size as in the data (an important point about simulation is that, once a particular CDR3 amino acid sequence has been chosen to not pass thymic selection, any future recurrence of that sequence in the simulation is also discarded); (ii) by deriving analytical mathematical expressions for the expected value (Methods). These predictions can then be directly compared to data.

\subsection{Model predicts many degrees of publicness in the data}\label{main_qfactor}
\begin{figure*}
\begin{center}
\textbf{A}
\includegraphics[width=0.4\linewidth,valign=t]{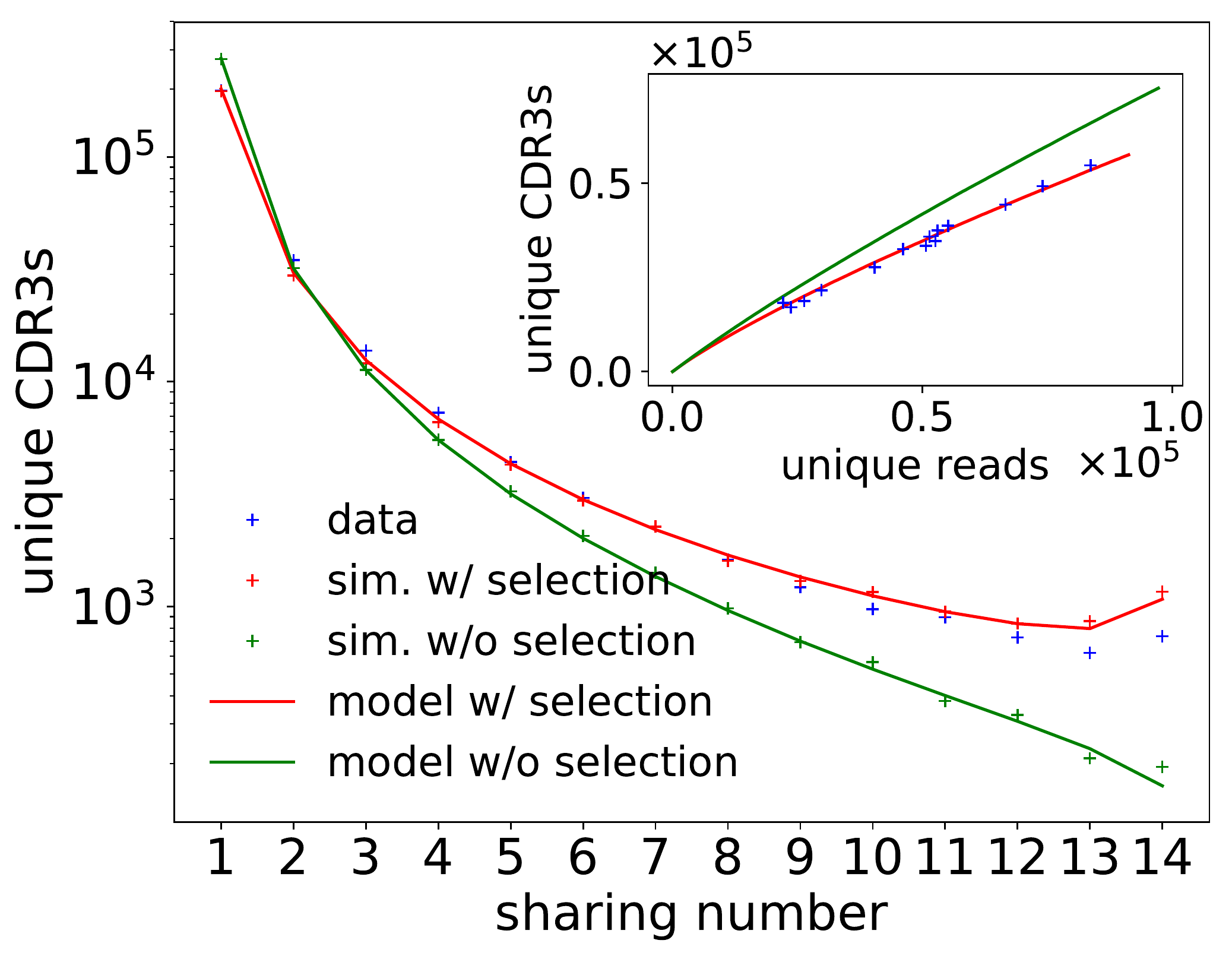}
\textbf{B}
\includegraphics[width=0.4\linewidth,valign=t]{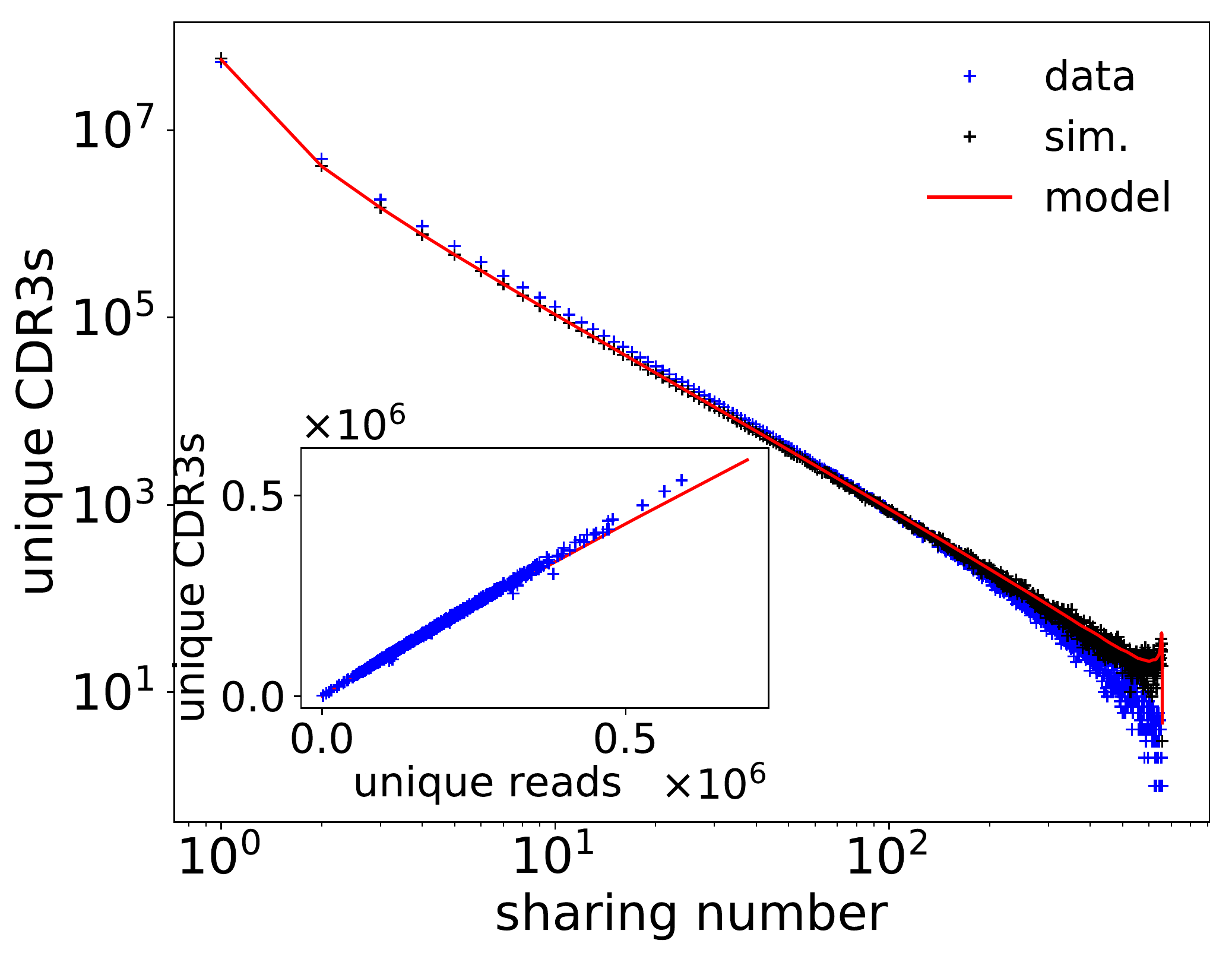}
\caption{Comparison of the distribution of shared sequences between model predictions with experimental samples. (A) Distribution of the number of sequences that are shared between $m$ individuals (sharing number on the x axis) for  14 mice. Data points (blue crosses) compared to  analytical model predictions (see section \ref{Methods_anal_sign}) with selection (red curves) and without selection (green curve) and simulations (see section \ref{Methods_sim_sharing}) based on the generation model  with selection (red crosses) and without selection (green crosses).  The model without selection underestimates sharing. The prediction is improved in the selection model. The model predictions derived from analytical calculations and stochastic simulations agree well (as they should).  The selection factor $q$, defined as the probability of a CDR3 to pass thymic selection, is inferred from a plot of the number of unique CDR3 amino acid sequences vs. the number of unique nucleotide sequence reads (inset, see Methods). (B) Comparison between the analytical (red line) and simulated (black crosses) model predictions and data (blue crosses) of the logarithm of the distribution of the number of sequences shared between $m$ individuals a cohort of 658 healthy humans. As for mice, the selection factor is determined by comparing the number of unique CDR3 amino acid sequences vs. the number of reads (inset).}
\label{sharing_nums_fig}
\end{center}
\end{figure*}

The comparison between data, model simulations and mathematical predictions shows excellent agreement in mice (Fig.~\ref{sharing_nums_fig}A) and humans (Fig.~\ref{sharing_nums_fig}B). The predictions depend on the only free parameter of the model, the selection factor $q$. This parameter was not set simply by fitting the sharing curves to the data. Instead, it was obtained independently as a proportionality factor required to explain the number of observed unique amino acid CDR3 sequences given the number of unique nucleotide sequences (insets of Figs.~\ref{sharing_nums_fig}A and B). This convergent recombination curve depends on $q$ in a predictable way (see Methods for mathematical expressions), making it possible to fit $q$ to the data (insets of Figs.~\ref{sharing_nums_fig} A and B). This method yielded selection factors of $q=0.15$ for mice, and $q=0.037$ for humans, surprisingly close to the estimate of $3\%$ for the fraction of human TCR that pass thymic selection \cite{Shortman1990}. Comparison of the prediction with and without selection in mice (red and green lines and points in Fig.~\ref{sharing_nums_fig}A) shows that adding selection greatly improves the agreement, despite a slight overestimation of high sharing numbers.

Humans have a much more diverse repertoire than mice \cite{Sethna2017}, resulting in lower number of shared amino acid TCR sequences. 
On the other hand, the very large cohort in the data set we analyze allows us to illustrate the very wide range of sharing behaviors. In particular, we find a long-tailed power-law distribution in the distribution of sharing numbers (Fig.~\ref{sharing_nums_fig}B), a feature that is reproduced by the model. A very small fraction of sequences are shared between all individuals in the $658$ donor cohort, while a large ($>90\%$) fraction of TCRs are found in just one sample. This diversity of behaviors reflects the diversity of generation probabilities implied by the strong biases in the VDJ recombination process that are correctly captured by our model.

\section{From samples to full repertoires}

\subsection{Sampling depth affects sharing}
\begin{figure*}
\begin{center}
\textbf{A}
\includegraphics[width=0.4\linewidth,valign=t]{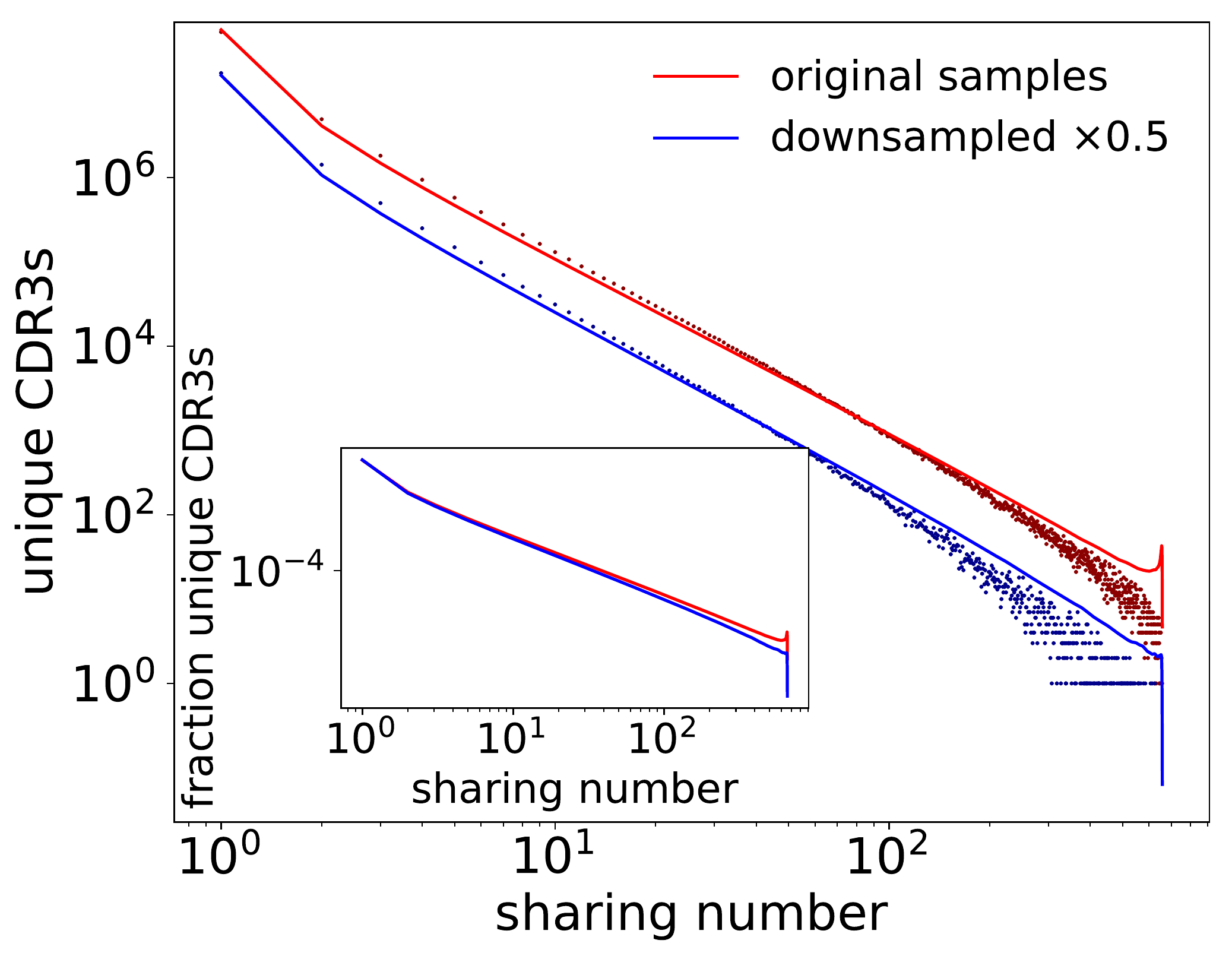}
\textbf{B}
\includegraphics[width=0.4\linewidth,valign=t]{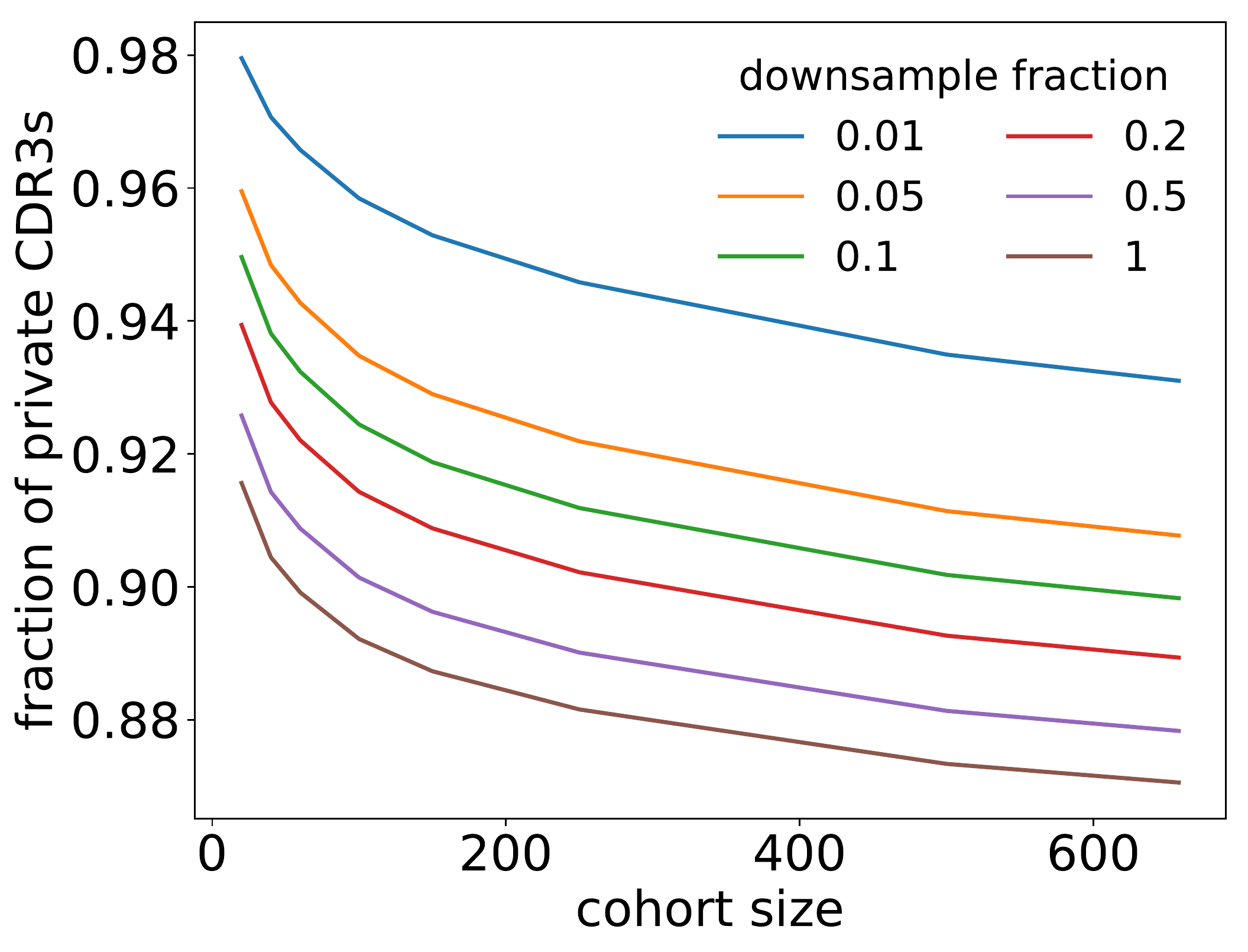}
\caption{The sharing number depends on the sampling depth and cohort study. Downsampling the number of sequences in all individuals affects sharing, and decreases the observed probability to be public. (A) The number of sequences shared as a function of sharing number decreases when the repertoires of all individual samples are downsampled by a factor 0.5 (blue  line) compared to the original sample (red line). The model predicts this change systematically (red and blue points). The distribution of the sharing numbers plotted using the fraction of CDR3s instead of absolute numbers (inset) demonstrates that the shape of the plot also changes. There are fewer highly shared sequences between individuals after downsampling. (B) The model prediction of the fraction of sequences that are all private (i.e. appearing in just one individual) as function of the downsampling factor and cohort size. Bigger samples or larger cohorts result in fewer private sequences, or more public ones, for any definition of public.}
\label{downsample_sharing_nums_fig}
\end{center}
\end{figure*}

While the sharing potential of a sequence depends just on its generation and selection probabilities, it is important to realize that actual sharing numbers will depend on the size of the cohort under study and the sampling depth of each individual T cell repertoire. To illustrate this  effect, we downsampled both the cohort size and the number of sequences in the human dataset, and recalculated sharing.
Fig.~\ref{downsample_sharing_nums_fig}A compares the distribution of sharing numbers in the original dataset, with the same distribution obtained from samples where a random half of the unique sequences were removed. The number of TCRs with each sharing number drops with downsampling, and this drop is more marked for high sharing numbers, as evidenced by the fraction of CDR3s with each sharing number (see inset of Fig.~\ref{downsample_sharing_nums_fig}A). In short, the more TCRs are captured in the repertoire samples, and the more likely sequences are to be shared. This effect is reproduced in detail by the model calculations.
This result generalizes previous observations that the number of shared TCRs between a pair of individuals should scale approximately with the product of the numbers of unique TCRs in each sample \cite{Murugan2012,Elhanati2014,Zvyagin2014a,Pogorelyy2017b} to arbitrary sharing numbers.

To demonstrate the effects of varying cohort and sample size more clearly,  we plot in Fig.~\ref{downsample_sharing_nums_fig}B the complementary quantity---the fraction of CDR3s which are purely \lq private\rq , i.e. present in only one repertoire.  This fraction decreases for large cohorts and large sample sizes. 
We note that cohort size and sample depth vary greatly from study to study; the data analyzed in this review go from a small cohort of mice (14 repertoires with a few tens of thousands TCRs each) to a very large cohort of humans (658 donors with 200,000 TCRs each). The strong dependence of the notion of privateness upon the parameters of the study cautions us against interpreting sharing numbers and public or private status of individual sequences too literally, and further emphasizes that publicness is not a binary but rather a continuous measure.  

\subsection{Cumulative diversity and extrapolation to full repertoires}
\begin{figure*}
\begin{center}
\textbf{A}
\includegraphics[width=0.3\linewidth,valign=t]{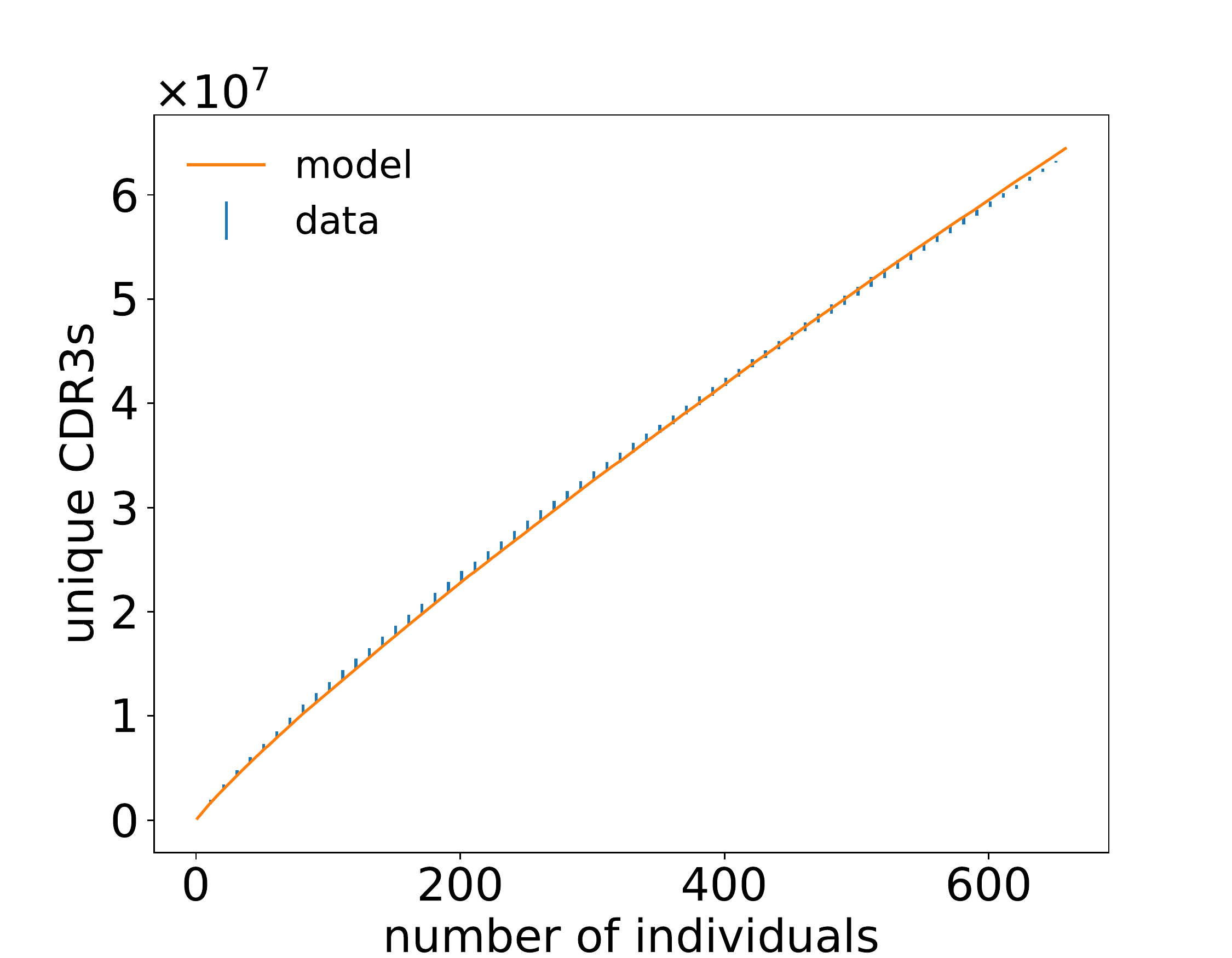}
\textbf{B}
\includegraphics[width=0.3\linewidth,valign=t]{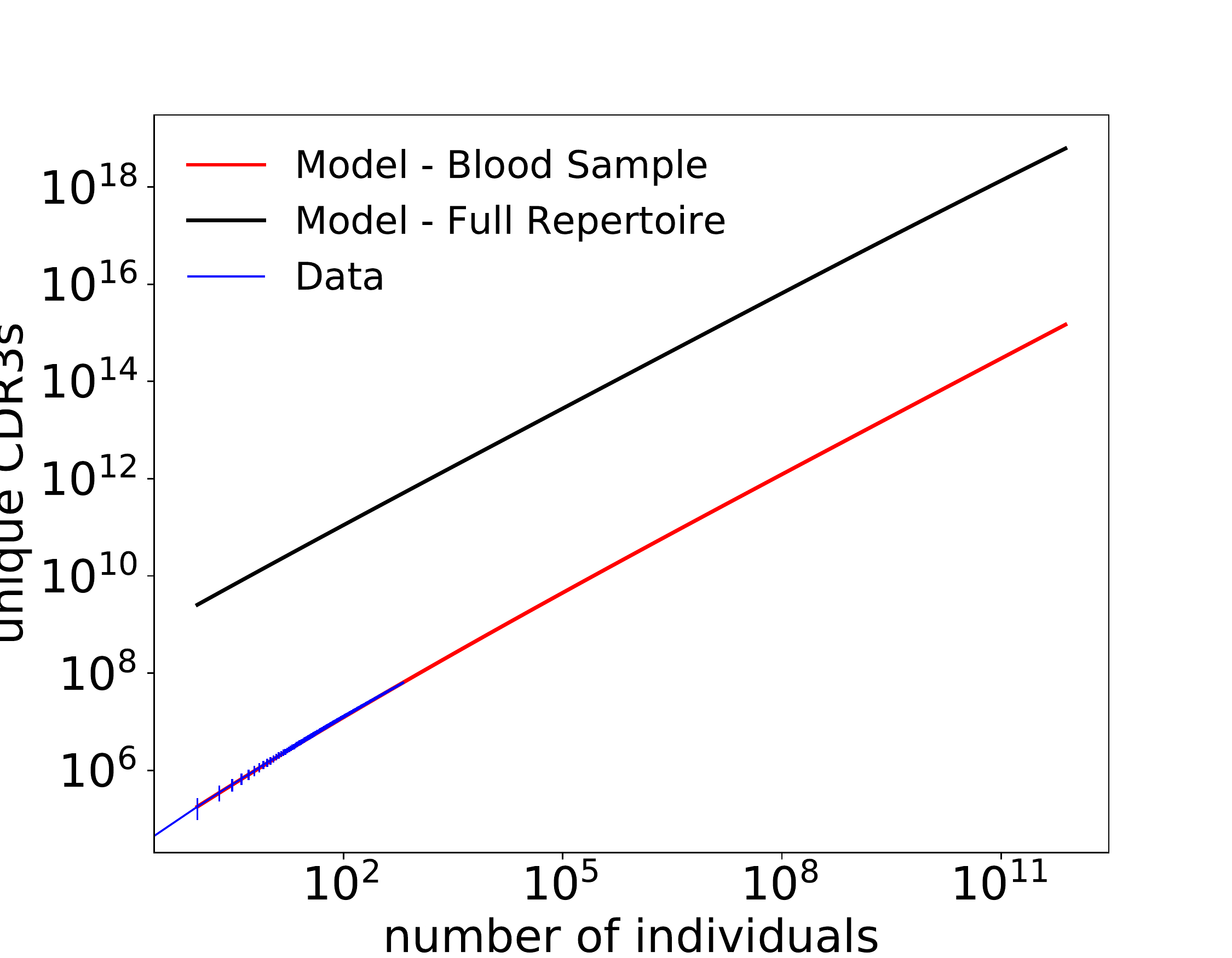}
\textbf{C}
\includegraphics[width=0.3\linewidth,valign=t]{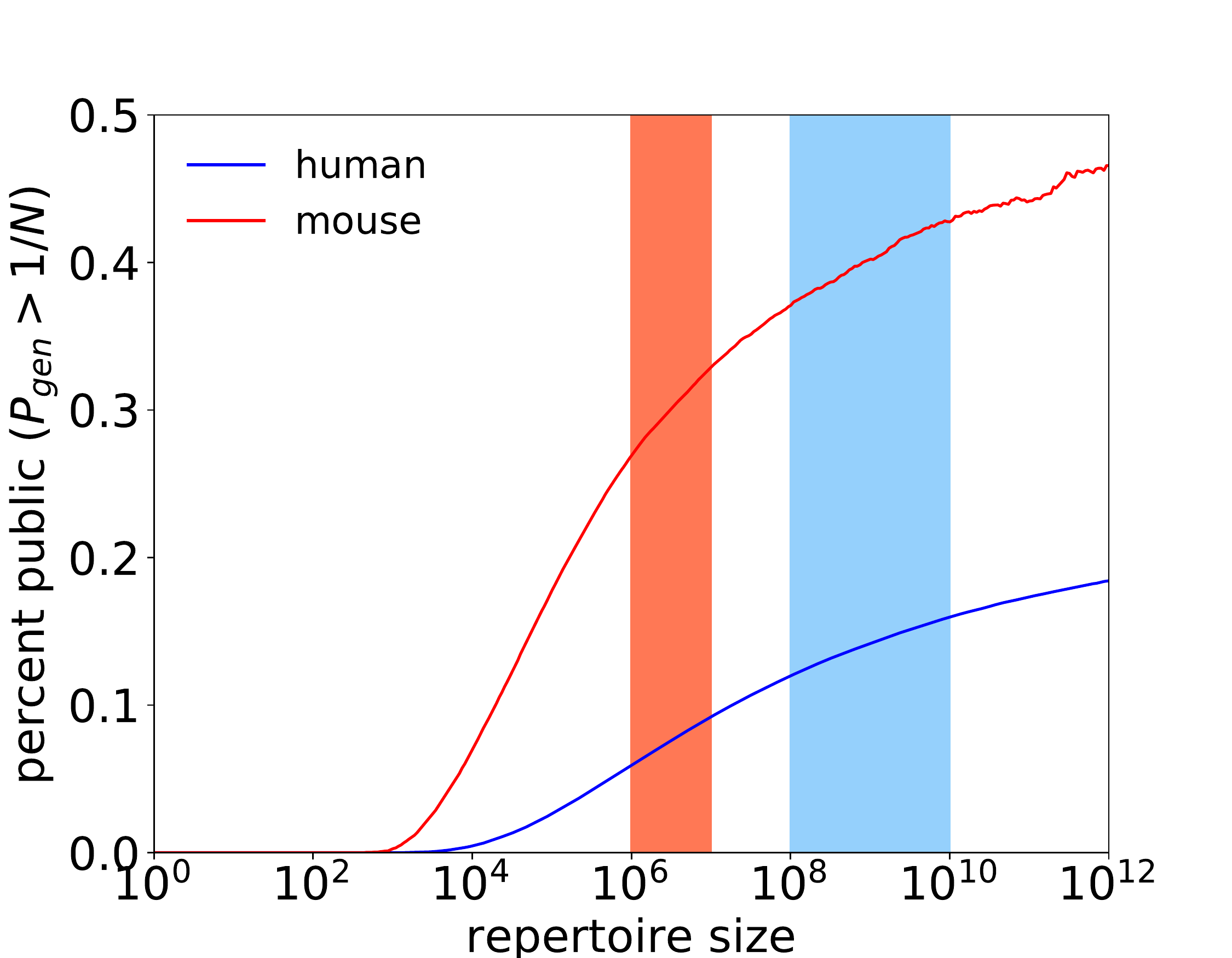}
\caption{The numbers of unique CDR3 amino acid sequences that are contained in the pooled repertoire of $n$ individuals, as a function of $n$. In principle, the rate of increase of the total number of cumulated sequences curve depends on the order of individuals in which the group is increased. In practice, for 30 random orderings of individuals, the numbers of cumulated unique CDR3 did not depend much on the order of adding new individuals, as can be seen from the very small error bars on the blue data points in (A). The theoretical prediction (red line, see Methods section \ref{methods_qfractor}) agrees very well with the data (blue points). The model prediction was obtained using the mean sample size of all 30 orderings. Each new individual adds $\sim 200,000$ new CDR3 sequences. (B) The observed cumulative number of sequence in $n$ individuals (blue line) compared to  the theory predication extended to very large cohorts (red line). This model prediction is based on an average sample size. The same prediction can be done for the full repertoires contained in the human body (with $10^{11}$ unique recombination events), which yields much larger numbers of unique CDR3s (black line). 
(C) Model prediction for the fraction of sequences in each individual that are truly `public', i.e. have a generation probability larger than $1/N$, where $N$ is the number of unique TCRs in each individual. The red and blue stripes mark the possible range of repertoire sizes in mice and humans, according to current knowledge. }
\label{adding_ind_fig}
\end{center}
\end{figure*}

As Fig.~\ref{downsample_sharing_nums_fig}B shows, 
most (more than $90\%$) amino acid TCRs are found in only one repertoire.
This means that, when pooling repertoires, each newly added repertoire will contribute a brand new set of TCRs to the pool.
To explore this idea, we define the \lq cumulative repertoire\rq\, obtained by pooling together the sampled repertoires of several individuals, and count the number of unique TCR$\beta$ amino acid sequences in it. This cumulative diversity grows almost linearly with the number of pooled samples (Fig.~\ref{adding_ind_fig}A), both in the data and according to the model (see Methods for calculation of the model prediction). The ratio of unique to total sequences starts at 1 for small numbers of pooled individuals, and decreases to around $0.9$ for high numbers of pooled individuals, consistent with the fraction of private sequences. 
It is interesting to ask whether this trend would continue for larger populations all the way up to the entire world population. Although we cannot answer this question directly by experiments, we can use the model to make predictions. Generating {\em in silico} repertoires for billions of individuals is of course impractical, but we can use mathematical expressions (Methods) to calculate the expected diversity.
Fig.~\ref{adding_ind_fig}B shows the theoretical cumulative diversity as a function of the number of individuals for up to $10^{12}$ individuals. Even with numbers of individuals largely exceeding the number of humans having ever lived ($10^{11}$), we are very far from saturating the space of observed TCRs.

The previous estimates rely on partial repertoires comprising a few hundred thousand unique TCRs obtained from small blood samples. However, the human body hosts $5\cdot 10^{11}$ T cells \cite{Jenkins2010}, and while the T cell population has a clonal structure, recent estimates of the number of clones, and thus of independent TCR recombination events, ranges from $10^8$ (from indirect sampling using potentially inaccurate statistical estimators \cite{Qi2014}), to $10^{10}$ (based on theoretical arguments \cite{Lythe2015}). The theoretical cumulative diversity based on that latter estimate of $10^{10}$ (Fig.~\ref{adding_ind_fig}B, black curve) still shows no sign of saturation. These results are a consequence of the enormous potential diversity of VDJ recombination, and indicate that the diversity of TCR$\beta$ is not exhausted even by the pooled repertoire of the entire world population.

Extrapolating these considerations to the full TCR repertoire of an individual allows us to estimate the fraction of truly `public' TCRs, defined as the sequences that are present in almost all individuals. If we define a public TCR sequence as one that has a generation probability larger than $1/N$, where $N$ is the number of T-cell clones in the body, then $1-e^{-1}=63\%$ of all individuals would be expected to have that sequence in their repertoire. With this definition, we can predict the percentage of public sequences as a function of repertoire size (Fig.~\ref{adding_ind_fig}C). Interestingly, this fraction ranges from 10 to 20\% for both humans and mice depending on estimates of the number of clones, despite their widely different TCR$\beta$ diversities and repertoire sizes. It is interesting to note that the lower diversity of the TCR$\beta$ repertoire in mice as compared to humans is matched in a proportional way to the ratio of the TCR repertoire sizes in the two species. 

\section{Predicting publicness}

\subsection{Sharing and TCR generation probability}
As we have seen, the sequence generation model correctly predicts the amount of sharing across individuals, as well as the fraction of public sequences. 
Underlying this prediction method is the idea that the likelihood that a given sequence will be shared is largely determined by the probability of generation of the sequence. Early versions of this argument \cite{Fazilleau2004,Venturi2008} noted that sequences with a high number of N insertions have lower generation probability (because of the diversity of possible insertions, each reducing the generation probability by a factor $\approx 1/4$), predicting that shared sequences would have fewer insertions than average. We have used recombination models inferred from data to refine this argument by accounting quantitatively for the effects of biases, convergent recombination, etc., on the probability of generation of particular TCR sequences. As a further test of the underlying ideas, we compute the generation probability of TCR sequences and ask how this quantity correlates with the sharing numbers.

\begin{figure*}
\begin{center}
\textbf{A}
\includegraphics[width=0.3\linewidth,valign=t]{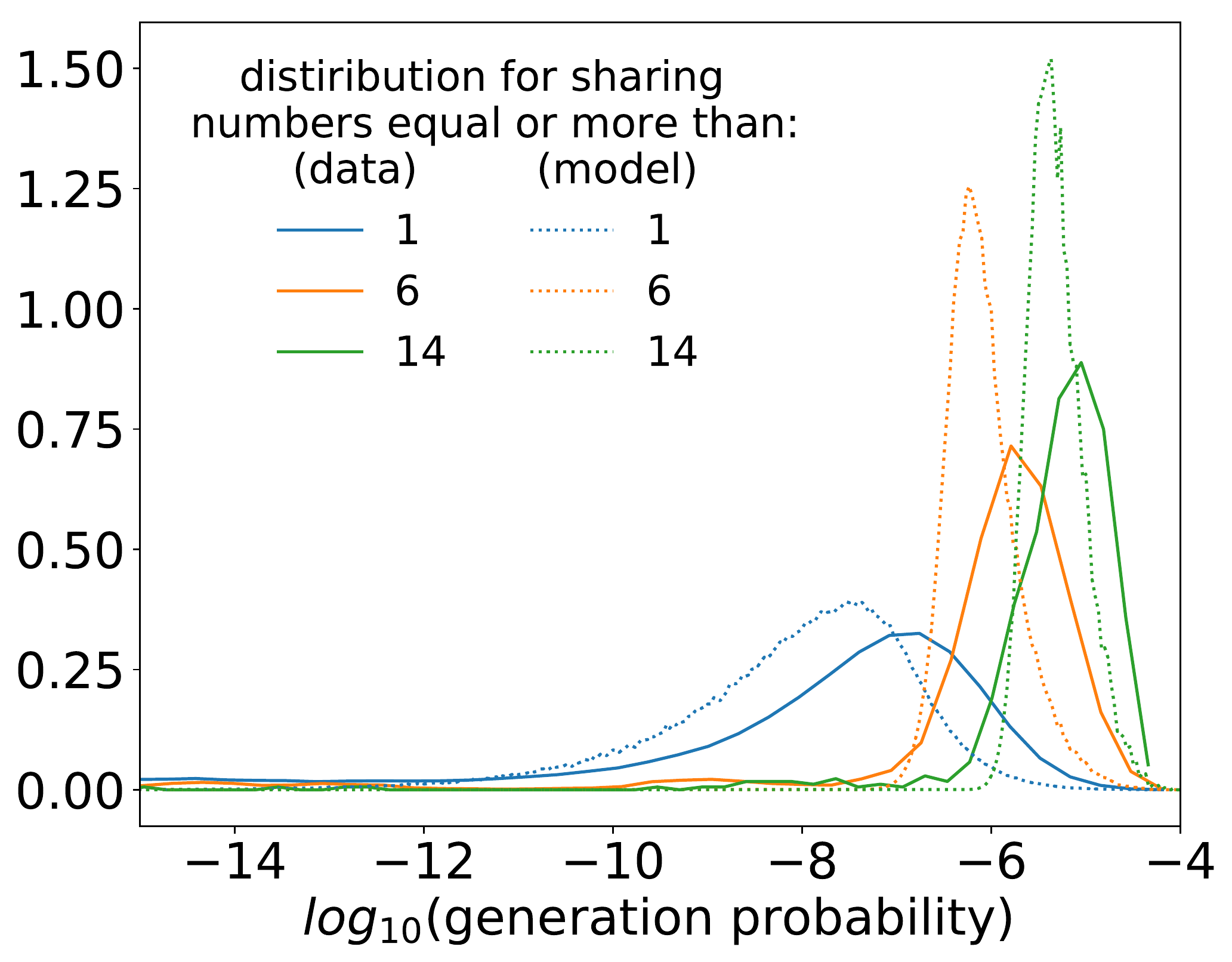}
\textbf{B}
\includegraphics[width=0.3\linewidth,valign=t]{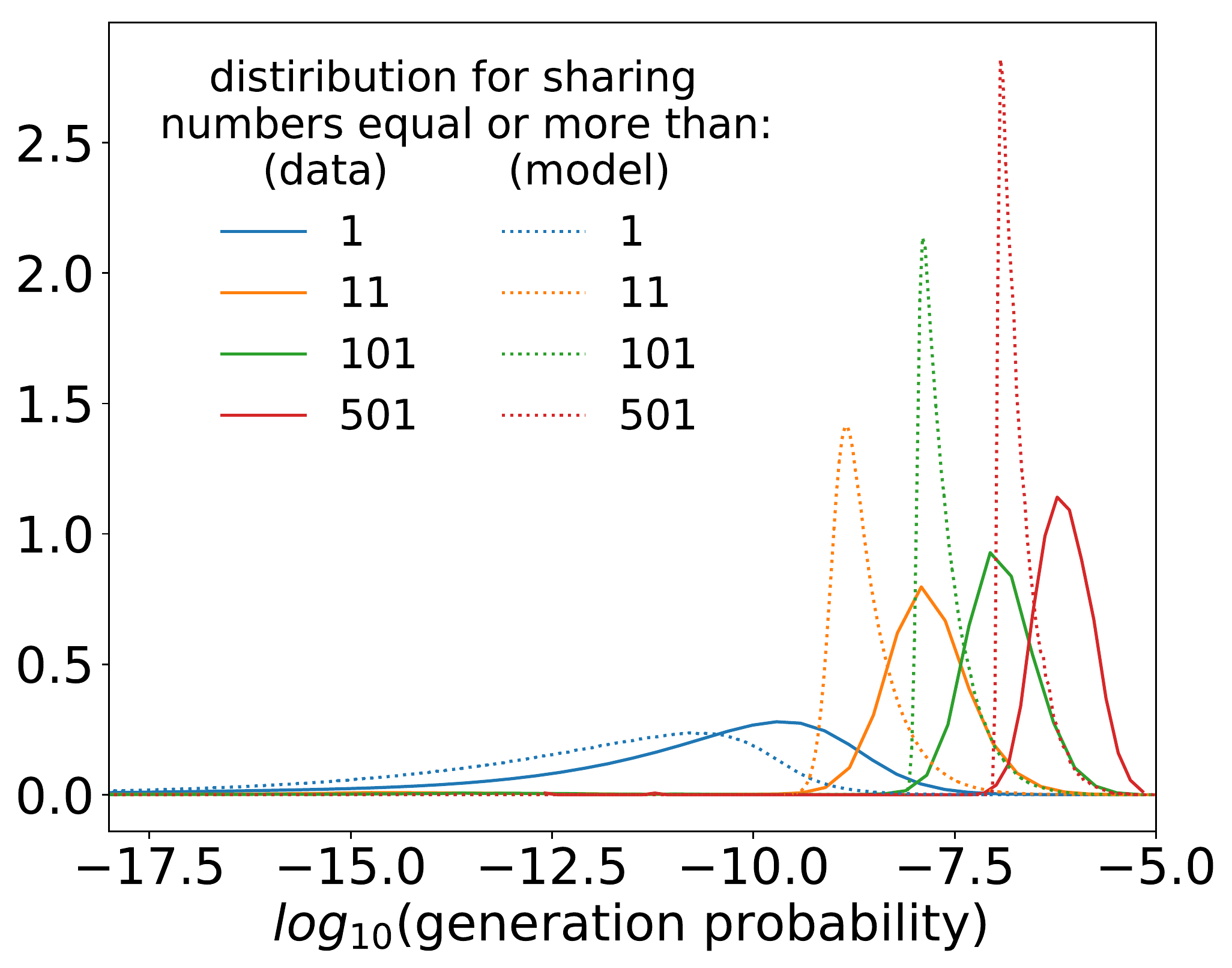}
\caption{Generation probability distributions for different minimal sharing numbers, for (A) mice (A) and (B) humans. For larger sharing numbers the distribution shifts toward higher probabilities and becomes narrower. This shift enables the characterization of the sharing number, or the degree of publicness, using the generation probability. The model captures the right trend of the sharing numbers, despite predicting much narrower distributions.}
\label{pgens_fig}
\end{center}
\end{figure*}

To calculate the generation probability of TCRs, one needs to sum the occurrence probabilities of all the possible recombination events leading to a given nucleotide sequence \cite{Murugan2012,Marcou2018} and, since we choose to follow CDR3 amino acid sequences, sum the probabilities of all nucleotide sequences leading to the amino acid sequence of interest. This is a computationally hard task that can be rendered tractable using a dynamic programming approach (see Methods).
We find that the distribution of generation probabilities of all TCR$\beta$ CDR3 amino acid sequences (Fig.~\ref{pgens_fig}, blue curves) is extremely broad, spanning many orders of magnitude. This observation is consistent with similar analyses at the level of nucleotide sequences in nonproductive \cite{Murugan2012} and productive \cite{Elhanati2014} human TCR$\beta$, in the $\alpha$ and $\beta$ chains of monozygous twins \cite{Pogorelyy2017a}, and mice \cite{Sethna2017}.
If we plot instead the generative probability distribution of sequences that are shared among two or more individuals in our data set, we find that the distribution narrows and shifts towards higher generation probabilities \cite{Murugan2012,Elhanati2014,Pogorelyy2017a} as expected. This effect is displayed in more detail in a plot of the generative probability distribution for sequences in our dataset with different sharing numbers (Fig.~\ref{pgens_fig}). On the same figure we plot the predictions of the recombination model, following the same protocol used for predicting sharing numbers (see Methods). There is a systematic shift between the predictions of the recombination model and the distribution of the data itself, for all sharing levels. This difference is due to the fact that the recombination model was inferred from non-productive sequences, and does not account for selection effects. The data sequences, however, have passed thymic and possibly other kinds of peripheral selection, affecting their statistics. The sequence-dependent nature of this effect was characterized and quantified in \cite{Elhanati2014}, with the general finding that  selection favors sequences with high generation probability. This is qualitatively consistent with the positive sign of the shift (solid lines versus dotted lines) we see in Fig.~\ref{pgens_fig}.
Our sharing calculations ignore any possible sequence dependence of selection, and instead selects TCRs at random (with probability $q$), regardless of their sequence identity.
The model prediction could in principle be improved by adding sequence-dependent selection factors to match the distributions as in \cite{Elhanati2014}. However, unlike the recombination model, such factors are expected to be specific to each individual, owing to their unique HLA type which is involved in thymic selection.

\subsection{PUBLIC: Classifier of public vs. private TCRs based on generation probability}

\begin{figure*}
\begin{center}
\includegraphics[width=.8\linewidth]{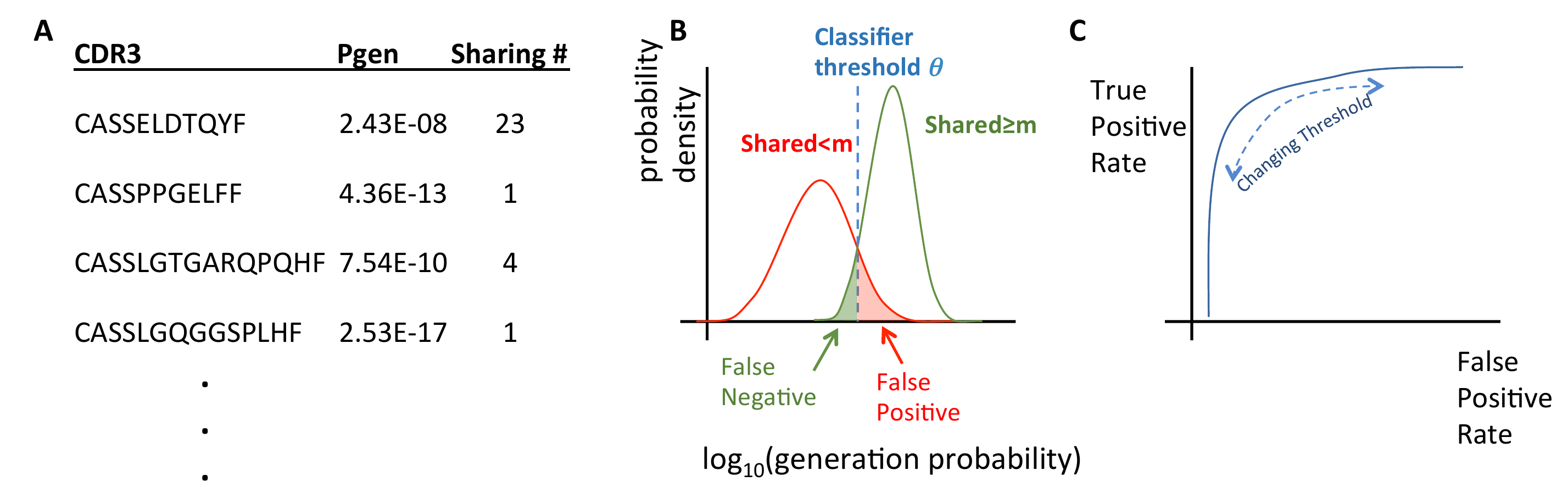}
\caption{A cartoon representation of the pipeline for the PUBLIC classifier. (A) Calculating the generation probability ($p_{\rm gen}$) and the sharing number for each CDR3 is the basis of the PUBLIC classifier. (B) The  $p_{\rm gen}$ distributions of shared sequences depend on the sharing number $m$. We compare the $p_{\rm gen}$ distributions of sequences shared between less than $m$ individuals to the distribution of sequences shared between more than $m$ individuals. We pick a classifier threshold value of $P_{\rm gen}$, $\theta$, that separates public from private sequences for this sharing number value of $m$. The areas of the histograms that fall on the wrong side of the threshold are defined as the false positive and false negative rates.  (C) For a given choice of the minimum sharing number $m$ we plot the true positive rate as a function of the classifier threshold $\theta$ to obtain a Receiver Operating Characteristic (ROC).}
\label{classifier}
\end{center}
\end{figure*}

The distributions of generation probabilities for the different sharing numbers suggest that the generation probability is a good proxy for the property of being {\it public}, regardless of the exact definition of publicness. We built a classifier called PUBLIC (Public Universal Binary Likelihood Inference Classifier), which is entirely based on the probability of generation computed as explained above (detailed in Methods) for each amino acid sequence (Fig.~\ref{classifier}A).
Before discussing the performance of this classifier, it is important to note that it is based on a model of recombination trained in a completely {\em unsupervised} way, {\em i.e.} without using any information about the public status of the sequences. In fact, this training can be done with IGoR \cite{Marcou2018} from the repertoire of a single individual, without including any sharing information. Unlike previous approaches \cite{Madi2014}, we do not fit additional model features based on the catalogue of sequences with their public or private status.

We arbitrarily define as \lq public\rq\, the TCRs that are found in at least $m$ repertoire samples among a total pool of $n$ individuals. The PUBLIC classifier calls a given TCR \lq public\rq\, if its generation probability is larger than a threshold $\theta$, calling it \lq private\rq\, otherwise. Intuitively, the threshold should be set to separate reliably the peaks in the probability density function of Fig.~\ref{pgens_fig} corresponding to different sharing numbers, as schematized in Fig.~\ref{classifier}B. The general performance of the PUBLIC classifier can be estimated by plotting the Receiver Operating Characteristic (ROC) curve, which represents the rate of false positives versus that of true positives as $\theta$ is varied (Fig.~\ref{classifier}C).

\begin{figure*}
\begin{center}
\textbf{A}
\includegraphics[width=0.4\linewidth,valign=t]{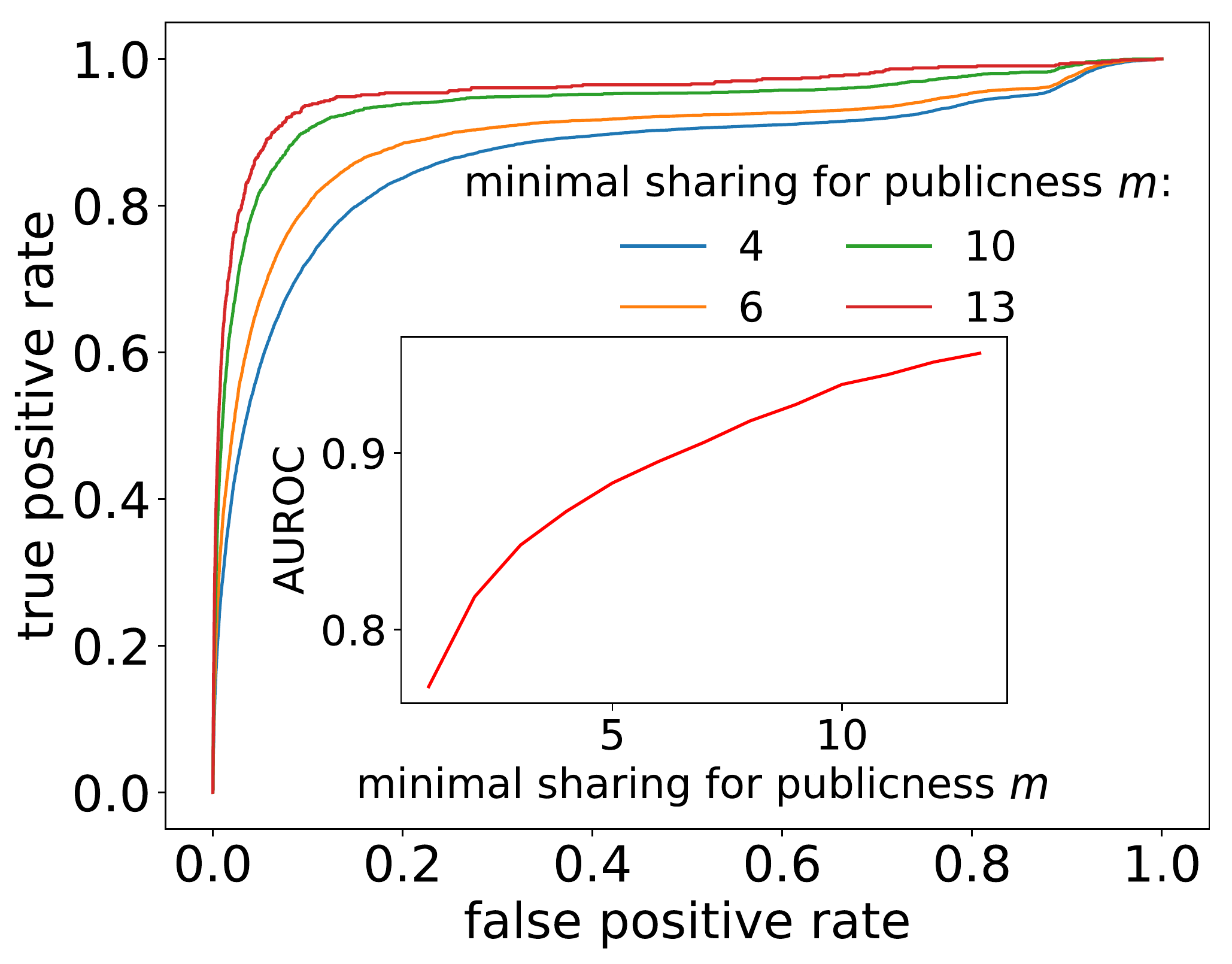}
\textbf{B}
\includegraphics[width=0.4\linewidth,valign=t]{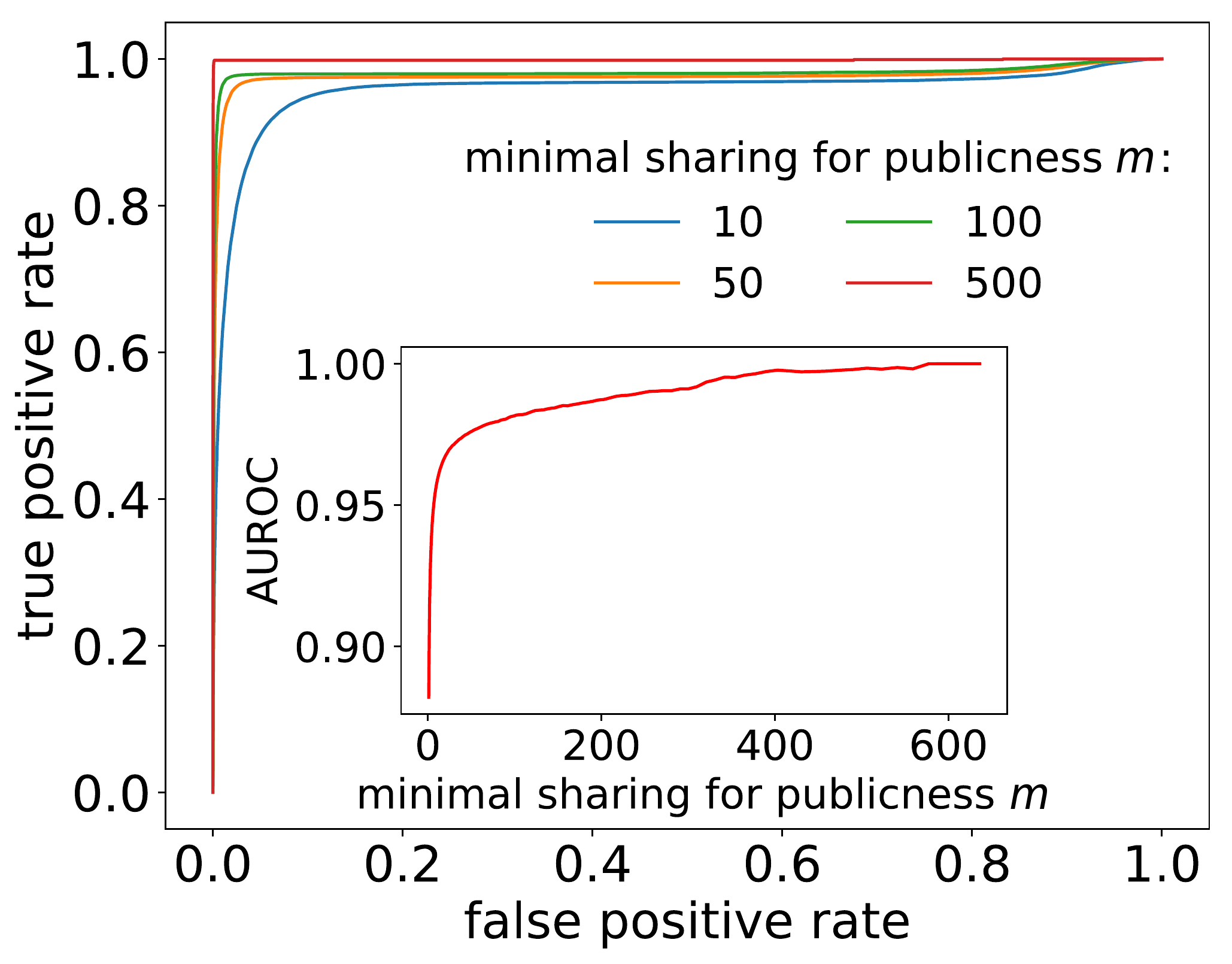}
\caption{Performance of the PUBLIC classifier. ROC curves for (A) mice and (B) humans for different minimal sharing numbers $m$. Inset: the Area Under the ROC Curve (AUROC) describes the probability of classifying a given sequence as public or private. Higher AUROC values correspond to a better a classifier. The AUROC score is increases with the minimal sharing number $m$ (inset), showing that a more restrictive definition of publicness gives better classifiers. }
\label{ROC_fig}
\end{center}
\end{figure*}

We plot ROC curves for a few different choices of $m$ (the minimal number of individuals with the TCR in their sampled repertoire for the sequence to be called public), for mice (Fig.~\ref{ROC_fig} A) and humans (Fig.~\ref{ROC_fig} B). The classification accuracy improves as publicness is defined to be more restrictive (larger $m$), although it performs well even for small $m$. 
For mice, the dataset we used had few individuals, making the operational definition of publicness less reliable.
However, for humans we find highly public TCRs are predicted almost perfectly by PUBLIC, despite the larger diversity of human TCRs. This suggests that the lesser performance of PUBLIC for mice may be attributed to the small size of the cohort, rather than to limitations of the classifier itself.

The performance of PUBLIC can be reduced to a single number by calculating the area under the ROC curve (AUROC). The AUROC corresponds to the probability that the classifier ranks a randomly chosen public sequence higher than a randomly chosen private one. The closer the AUROC score is to 1, the better the classifier.
As was clear from the ROC curves themselves, the AUC improves as the degree of publicness is higher (insets of Fig.~\ref{ROC_fig}A-B). As the minimal sharing number $m$ increases, the classifying task becomes easier and the prediction better. In fact, having the minimal sharing number $m$ close to the cohort size $n$ will in general make publicness rarer, and the public sequences more extreme in their generation probabilities. 

\section{Public specific response}
Sharing properties are interesting in their own right, but they also provide a basal expectation for the prevalence of certain TCRs. Using the sharing prediction, one can identify TCRs that are more shared in specific populations or subsets than expected according to the recombination model. When counting sharing in a population of individuals affected by a common condition, this `over-sharing' can be indicative of a specific T-cell response to the antigens associated with the condition. Such sharing of specific TCRs is expected from the relatively low diversity of antigen-specific sequences revealed by {\em in vitro} multimer-staining experiments \cite{Dash2017,Glanville2017}.
A very similar idea has been exploited by several groups to identify TCRs specific to the Cytomegalovirus  \cite{Emerson2017}, Type-1 diabetes  \cite{Seay2016,Fuchs2017}, arthritis \cite{Faham2017} and other immune diseases  \cite{Zhao2016}. In these studies, there is no theoretical expectation from the recombination model. Rather, the basal expectation for TCR sharing is given by a negative-control cohort. However this control can be efficiently replaced by the recombination model presented here, as demonstrated in \cite{Pogorelyy2017}. In this analysis, specific TCRs emerge as outliers that are shared much more frequently than predicted by the model.

We wondered whether such an approach could be useful for identifying tumor-specific TCRs as sharing outliers among cancer patients. The T-cell repertoire of tumor-infiltrating cells has been studied to look for signatures of immunogenicity \cite{Li2016,Pasetto2016,Snyder2017}, and the overlap between the tumor and blood repertoires was shown to predict survival in glioblastoma patients \cite{Hsu2016}. In addition, the tumor-specific TCRs have been reported to be shared in the tumor-infiltrating and blood T-cell repertoires of breast cancer \cite{Munson2016}.

\begin{figure}
\begin{center}
\includegraphics[width=0.8\linewidth,valign=t]{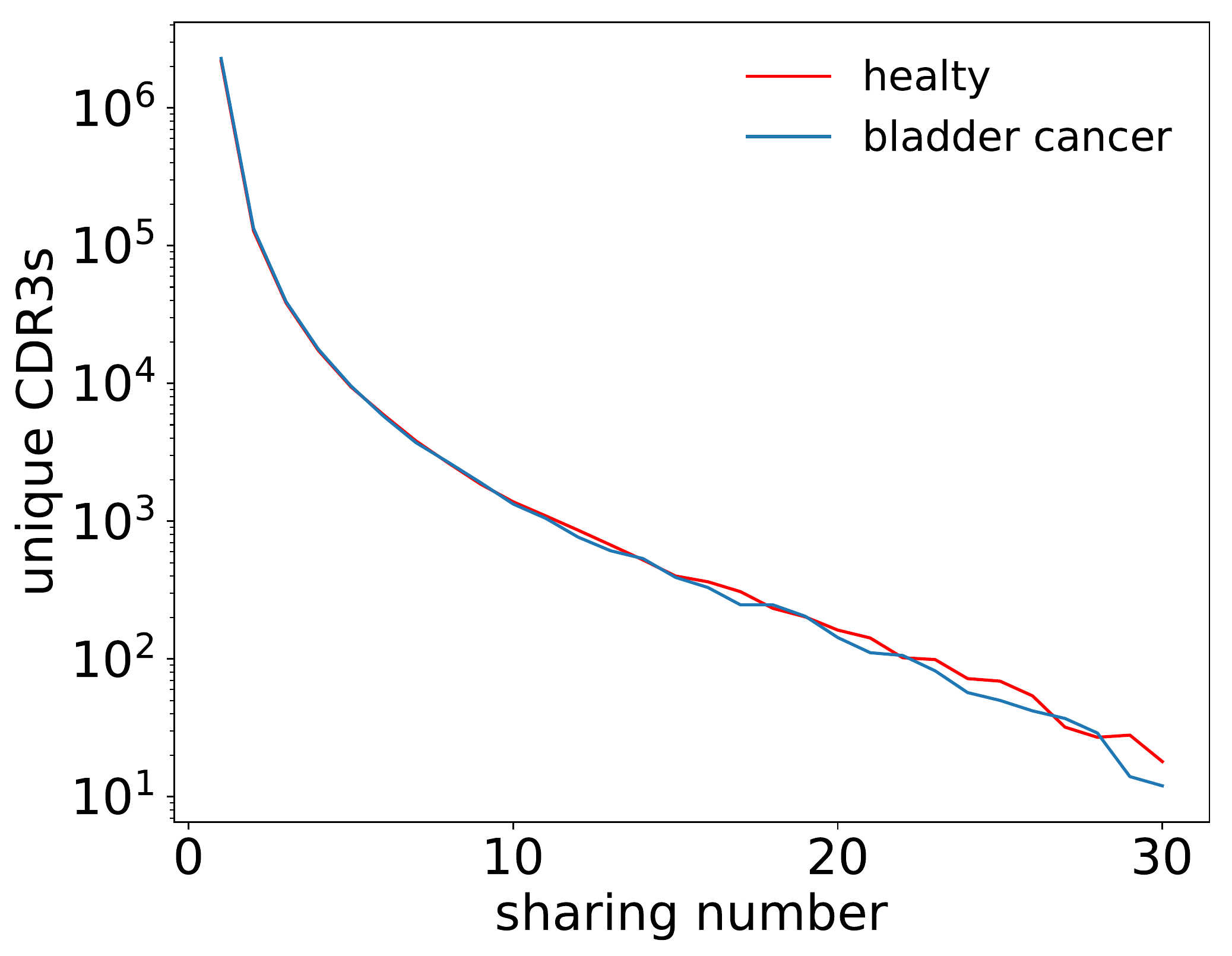}
\caption{The sharing of sequences between bladder cancer patients. The distribution of sharing numbers calculated based on the repertoires of 30 bladder cancer patients compared to 30 healthy individuals downsampled to have the same sample sizes. The distribution of sharing numbers is the same in healthy and bladder cancer patients, indicating that there are no statistically significantly over-represented TCRs in the blood repertoire of cancer patients.}
\label{sharing_cancer}
\end{center}
\end{figure}

We thus asked whether the blood repertoires of patients with bladder cancer contained TCRs with more sharing than would be predicted by our recombination model. We performed the sharing analysis on 30 patients with bladder cancer, on TCR repertoires sequenced from blood samples \cite{Snyder2017}. We compared it with 30 healthy individuals, chosen at random among the individuals studied in Ref.~\cite{Emerson2017} to have similar sample sizes. We then downsampled the reference  repertoires of the healthy individuals to have the exact same sample sizes as the cancer patients to guarantee a fair comparison. We found that the numbers of shared sequences in the blood of bladder cancer patients are almost identical to those found in the healthy samples, and thus also in agreement with the recombination model (Fig.~\ref{sharing_cancer}). This is consistent with previous reports that did not find any signatures of TCR repertoire anomalies in the blood of bladder cancer patients, although some small differences could be seen in the tumors. There are many possible explanations for this observation: the tumor-specific response may be statistically negligible amid the large number of other cells; or the response may not have propagated to the blood; or different patients generate responses against different neoantigens; or they generate very different responses against the same neoantigen; or the tumor does not generate any response at all. Tumor samples from larger cohorts would be needed to distinguish between these different hypotheses. Additionally this result is only true for bladder cancer. Different tumor types that have a higher rate of infiltration to the blood may be more likely to result in detectable signatures in the blood.

\section{Discussion}
In this paper we extensively tested and quantified the previously proposed hypothesis \cite{Venturi2008,Venturi2013} that public TCRs owe their status to the ease of generating them through V(D)J recombination. Predicting and characterizing TCR sharing and publicness is important to identify universal features of the immune response across individuals. This knowledge can be useful when designing vaccines that have a high probability of eliciting an immune response, or for identifying candidate T-cell clones in immunotherapeutic strategies \cite{Sadelain2017}.

Our predictions, and their agreement with the data, support the notion that `publicness', as it is usually defined, is context-dependent \cite{Venturi2008}.
The public status of a TCR depends not only on its (intrinsic) generation probability, but also on (extrinsic) parameters including the number of individuals sampled, the sequencing depth of the samples, and the definition of publicness -- the minimal number of individuals that need to share a TCR to call that TCR public.
Instead, we have showed that we can define the potential for publicness, largely determined by the generation probability of the sequence, and use it to predict actual sharing numbers for any set of repertoire samples. At the same time, we proposed that an absolute notion of publicness can be defined based on the full repertoire of individuals. According to this definition, a TCR is public if its probability of occurrence is larger than the inverse of the number of unique TCRs hosted in the entire repertoire. While this definition is impossible to explore directly in humans, for whom only repertoire samples can be obtained, our data-driven recombination model can make predictions about the public status of particular sequences, and the fraction of the repertoire that is public, using this specific definition (Fig.~\ref{adding_ind_fig}).

We report a wide spectrum of publicness, which we show arises from the very wide distribution of TCR generation probabilities. The high-end of the distribution holds sequences that will be included in any healthy repertoire, just by virtue of their high generation probability. Due to their publicness, it had been conjectured that some of these common TCRs might have a close to innate function \cite{Venturi2013}. In this context it should be noted that young, pre-birth repertoires are known to be much less diverse both in humans \cite{Pogorelyy2017a} and mice \cite{Sethna2017}, due the late appearence of TdT, the enzyme responsible for insertions in the recombination process. Consequently, the pre-birth repertoire is expected to be much more public that the adult one, and could be enriched in innate-like TCRs. However, since no conclusive evidence has been provided about the functional role of these high probability sequences, we cannot rule out the possibility that they are just there by chance, without a specific function. The other end of the TCR distribution---the long tail of low generation probabilities---contributes to the private part of the repertoire, which makes up the majority of the repertoire according to our estimates. It would be interesting to explore whether these sequences have a functional role or are just by-products of the recombination process. 

High-throughput TCR repertoire datasets contain abundance levels (number of reads) for each TCR. TCR abundances have be attributed to convergent recombination, implying a correlation between high abundance and publicness \cite{Venturi2008}. However, this connection may be confounded by other processes affecting the abundance levels reported by high-throughput sequencing. A big source of diversity in TCRs abundances is the peripheral proliferation of some TCRs, regardless of their generation probability. In addition, experimental or phenotypic noise, including PCR amplification noise \cite{Best2015b} and expression variability (for cDNA sequencing) also affect reported abundances. These various effects are expected to dilute the correlation between abundance and publicness. Note that our statistical models are constructed based only on unique sequences, circumventing clonal expansion dynamics, and ignoring abundance levels altogether.

The sharing analysis naturally leads to defining the PUBLIC score, which we show predicts sharing properties with high accuracy.
The PUBLIC score is learned in an unsupervised manner, using a statistical model trained with no information about the sharing status of TCRs. Thus, sharing can be very well predicted with neither abundance nor sharing information. This success suggests that being public is a very basic property of the recombination process itself, and also provides a strong validation of the recombination model. It would be interesting to explore how using TCR sharing status and abundance levels in a supervised manner that refines the classifier could lead to better predictions.

Our prediction for sharing is mainly based on the generation model \cite{Marcou2018}, which is sequence specific, attributing each sequence its own probability of generation. We have found that an overall selection factor is needed to predict sharing numbers correctly, but this simple and effective model is sequence independent. Previous work \cite{Elhanati2014} inferred a sequence-specific selection process by comparing generation model results to observed sequences. In principle such a model could be combined within our framework to yield refined sharing predictions. While the parameters of the generation process are largely invariant across individuals \cite{Marcou2018}, selection is expected to be individual-dependent and heritable due to the diversity of HLA types in the population \cite{Rubelt2016}. The large variability in the V and J genes selection pressures inferred in \cite{Elhanati2014} is consistent with this notion, but in the same work some amino-acid features of selection were found to be universal. Quantifying these universal features and including them in the model could both improve the predictions for the sharing numbers, and enable a better assessment of the potential publicness of specific sequences through an improved classifier.

The discussion in this work was focused on TCR$\beta$ chains, but in general can be applied to any recombined chain, including $\alpha$, $\gamma$ and $\delta$ TCR chains, as well as B-cell receptor (BCR) light and heavy chains, or to paired chain combinations. The $\alpha$ chain, as part of the $\alpha\beta$ receptor, contributes to antigen recognition. It is less diverse than the $\beta$ chain, implying higher sharing numbers \cite{Pogorelyy2017a}. Paired $\alpha\beta$ data is becoming available as paired sequencing technologies improve \cite{Howie2015,Grigaityte2017}, but the resulting repertoires are currently too small or not yet available for analysis. As more paired sequencing data becomes available, it will be interesting to study the sharing properties of the $\alpha\beta$ repertoire using recombination models for pairs.

A similar analysis could be performed on BCRs. The problem is further complicated by somatic hypermutations, which add further diversity and are expected to reduce sharing as well as the ability to predict it. However, the role of the generation probability, for which the models have been trained \cite{Elhanati2015,Ralph2016,Marcou2018}, for sharing and publicness has not been explored. Machine learning approaches to predict publicness of BCR \cite{Greiff2017} could be combined with estimates of the probability of generation and hypermutations profile \cite{Yaari2015,Mccoy2015,Marcou2018} to provide accurate predictions for the public status of BCRs. Such an analysis applied to the result of affinity maturation in different individuals infected with the same pathogens \cite{Georgiou2014} could be used to assess the impact of convergent recombination in the public response and better understand the evolution of specific antibodies, and guide vaccination strategies to facilitate the emergence of broadly neutralizing antibodies \cite{Corti2013}.

\section{Methods}

\subsection{The probability of generating a TCR sequence}
To evaluate TCR generation probabilities, we first constructed a probabilistic generation model of the recombination process \cite{Murugan2012}. Such a model is parametrized by probabilities for each choice of V,D and J gene, for each deletion length of the different genes, and for each insertion pattern of random nucleotide between the genes. Then the probability of a recombination scenario is 
\beqn
P_{\rm sc}=&P({\rm V,D,J})P({\rm del}V|{\rm V})P({\rm delD}|{|\rm D})P({\rm delJ|J}) \nonumber \\
& P({\rm insVD})P({\rm insDJ})
\eeqn
The probability of a TCR sequence, whether it is a nucleotide of amino-acid sequence, for the full sequence or just the CDR3, is obtained by summing the above probability over all the possible scenarios leading to the sequence of interest.

The generation model can be inferred efficiently using the IGoR software \cite{Marcou2018} from non-functional recombinations, which produce out-of-frame or stop codon containing sequences. Model training is done by finding model parameters that maximize the likelihood of the data, equal to the product of generation probabilities of the observed TCRs in the dataset. \YE{Here we used IGoR to infer a generation model from the non-functional reads in the datasets from which the productive reads used for the sharing analysis came, human data in \cite{Emerson2017}, and mice data in \cite{Madi2014}}.

To calculate the generation probabilities of CDR3 amino-acid sequences, we used an efficient algorithm that avoids brute-force summation of all possible scenarios using dynamic programing.

\subsection{Evaluating the number of shared sequences using simulations}\label{Methods_sim_sharing}

Once inferred, a generative model can be used to generate random {\em in silico} samples of any size. Recombination scenarios are generated using Monte Carlo sampling by drawing events such as gene choices, deletions and insertions according to the model parameters. Each recombination scenario
constructs a nucleotide sequence which is filtered for productivity (in-frame, no stop codons or pseudogenes, and the conserved residues C and F are present). A productive nucleotide sequence is then trimmed to the CDR3$\beta$ region and translated into an amino acid sequence. To model thymic selection only a random fraction $q$ of the productive CDR3$\beta$ sequences are considered. This is implemented using a hash function, keeping only sequences whose normalized hash values are less than $q$. This negative selection process is a random function of the sequence, which is consistent between any simulated individual sample, so that a given CDR3$\beta$ will either pass or fail selection in all individuals. A simulated sample can thus be generated to match the cohort size and sequencing depth of the real data, and then analysed with with the same pipelines.

\subsection{Analytical calculation of the number of shared sequences}
\label{analytics}

\subsubsection{Predicting sharing numbers from the distribution of generation probabilities}\label{Methods_anal_sign}
Given a collection of CDR3$\beta$ sequences $s \in S$, a model that assigns probabilities $p(s)$ for each sequence, and $N$ independent sequences drawn from the model, the expected number of observed unique sequences $M_0$ is:
\beq
M_0(N) = \sum_{s\in S} 1 - (1-p(s))^N \approx \sum_{s\in S} 1 - e^{-p(s)N}
\label{eq_Mo}
\eeq
where we have made the Poisson approximation for small $p(s)$. If there are $n$ individuals, with sample sizes $\{N_i\}$, then the expected number of sequences which will be found in exactly $m$ individuals (sharing number $m$) is:
\beq
M_m(\{N_i\}) = \sum_{s\in S} \sum_{J \in J_m} \prod_{j \in J} \left(1 - e^{-p(s)N_j} \right) \prod_{i \notin J} e^{-p(s)N_i}
\label{eq_Mk}
\eeq
where $J_m$ is the collection of all possible combinations of m individuals. This can be computed more efficiently by use of the generating function $G(x , \{N_i\})$:
\beqn
G(x , \{N_i\}) & = &\sum_{m = 0}^{n} M_m(N_i) x^m \nonumber \\ 
& = &\sum_{s \in S} \prod_{i =1}^n \left[ e^{-p(s)N_i} + \left(1 - e^{-p(s)N_i} \right) x \right]
\label{eq_G}
\eeqn
where the $M_m$s are the coefficients of the polynomial $G(x, \{N_i\})$, and can be calculated just by expanding the polynomial in $x$ and summing over $s$.

\subsubsection{Density of states approximation}
While the above equations are exact, summing over each individual sequence is intractable given the huge number of sequences. Instead, an integral approximation based on the ``density of mstates'' is used. Let us call $E(s)=-\ln p(s)$ the Shannon surprise of generating sequence $s$ at random, also formally equivalent to an energy in physics according to Boltzman's law.
The density of states, $g(E)dE$, counts the number of sequences between $E$ and $E+dE$. Summation of an arbitrary function $\Phi(p(s)) = \Phi(E(s))$ over the states (sequences) in $S$ can then be turned into an integral:
\beq
\sum_{s\in S} \Phi(p(s)) \approx \int_{0}^{+\infty} g(E) dE \Phi(E)
\eeq
A numerical estimation of $g(E)$ is required to compute this integral. Estimating $g(E)$ is done by drawing large Monte Carlo samples of sequences ($10^7$ for humans and $10^6$ for mice) from the generative model and calculating the generation probabilities of each sequence. Values of $E(s)=-\ln p(s)$ can then be histogrammed into bins of size $dE$ and the resulting distribution normalized to integrate to 1. This yields a probability density, $P(E)$ (shown Fig.~\ref{pgens_fig}), which can be used to compute the density of states:
\beq
\begin{split}
P(E)dE & \approx \sum_{s\, |\, E<-\ln p(s)<E+dE} p(s) \approx g(E)e^{-E}dE\\
g(E) &\approx P(E)e^{E}.
\end{split}
 \eeq 
Equations \ref{eq_Mo} and \ref{eq_G} can now be rewritten in terms of integrals:
\beq
M_0(N) \approx \int_0^{+\infty} g(E)dE \left(1- e^{-N e^{-E}}\right),
\label{eq_Mo_approx}
\eeq
and
\beq
G(x , \{N_i\})  \approx \int_0^{+\infty} g(E)dE \prod_{i=1}^n \left[ e^{-N_i e^{-E}} + \left(1 - e^{-N_i e^{-E}} \right) x \right].
\label{eq_G_approx}
\eeq

\subsubsection{Sharing modified by selection}
While the above analysis is general, it depends on the state or sequence space $S$ (the collection of productive CDR3$\beta$s that pass selection) and on a model that assigns probabilities to each sequence. The preferred model to use will be the probability of generating a sequence ($p_{\rm gen}(s)$), however this model is defined and normalized over a state space of all possible recombination events, many of which lead to non-functional or negatively selected sequences. As a result, the model $p(s)$ that will be used needs to be renormalized to reflect the reduced sequence space of productive, selected sequences. This introduces two factors. First factor, $f$, is the fraction of sequences which are functional (in-frame, no stop codons or pseudogenes, conserved residues are present), and can be computed directly from the generative model ($f = 0.236$ for humans and $f = 0.260$ for mice). The second factor, $q$, is the fraction of productive sequences which pass selection and must be inferred (see below). These two factors provide the definition for the model that is used in the analysis:
\beq
p(s)=\frac{p_{\rm gen}(s)}{fq}
\eeq
The effect of renormalizing $p_{\rm gen}(s)$ to $p(s)$ on the density of states is that the energies are shifted by a constant $\ln f + \ln q$ and is everywhere reduced by a factor of $f \times q$:
\beq\label{gE}
g(E) = g_{\rm gen}^*(E - \ln f - \ln q)\times f \times q
\eeq
Where $g_{\rm gen}(E)$ is the density of states computed from $p_{\rm gen}(s)$ and $g(E)$ is derived from $p(s)$. 

\subsubsection{Inferring the selection factor $q$}\label{methods_qfractor}
The selection factor $q$ is inferred by running a least-squares regression on the model predictions for the $M_0(N)$ curve (Eq. \ref{eq_Mo_approx}). This curve relates the number $M_0$ of unique amino acid CDR3 sequences observed to the number $N$ of productive, selected recombinations generated. To determine the $M_0(N)$ curve from the data, the number of productive selected recombinations must be determined for each sample. Fortunately, due to the limited sequencing depth, the number of unique productive nucleotide reads in each individual sample is very close to the actual number of selected recombinations. In practice, $N$ was taken to be the number of unique nucleotide sequences of each repertoire, summed over a subset of the individuals, and $M_0$ was the number of unique amino-acid sequences resulting from the translation of the aggregated repertoire of the same subset of individuals. The curve was obtained by adding more and more random individuals to the subset, and averaged over 30 realizations of that random addition process (Fig. \ref{adding_ind_fig}A).
A least-squared regression of Eq.~\ref{eq_Mo_approx} with Eq.~\ref{gE} to that empirical curve yielded a value for $q$ of approximately 0.037 for humans and 0.155 for mice.

\subsubsection{Analytic computation of public fraction of a repertoire}\label{Methods_public_anal}

In Fig.~\ref{adding_ind_fig}C  a sequence $s$ in a repertoire of size $N$  is defined as public if $p(s) \geq 1/N$. The fraction of the repertoire comprised of these sequences is computed by evaluating:
\beq
\textrm{fraction public}=\frac{\int_0^{\ln(N)} g(E)dE \left(1- e^{-N e^{-E}}\right)}{\int_0^{+\infty} g(E)dE \left(1- e^{-N e^{-E}}\right)},
\eeq
where the term in parenthesis corresponds to the probability that a given sequence with probability $e^{-E}$ is found in a repertoire of size $N$.


{\bf Acknowledgements.} The work of TM and AMW was supported in
part by grant ERCCOG n. 724208. The work of ZS and CC was supported in
part by NSF grant PHY-1607612. The work of YE was supported by a fellowship
from the V Foundation. The work of CC was performed in part at the Aspen Center 
for Physics, which is supported by National Science Foundation grant PHY-1607611.
\medskip

\bibliographystyle{pnas}

\end{document}